\documentclass[universe,article,accept,pdftex,moreauthors]{Definitions/mdpi}

\firstpage{1} 
\pubvolume{1}
\issuenum{1}
\articlenumber{0}
\pubyear{2024}
\copyrightyear{2024}
\externaleditor{Academic Editor: }
\datereceived{ } 
\daterevised{ } 
\dateaccepted{ } 
\datepublished{ } 
\hreflink{https://doi.org/} 

\usepackage[english]{babel}
\usepackage{amsfonts}
\usepackage{comment}

\Title{Scaling Behaviour of $dN/dy$ in High-Energy Collisions} 

\TitleCitation{Scaling Behaviour of $dN/dy$ in High-Energy Collisions} 

\Author{G{\'a}bor Kasza 
 $^{1,2,}$* \orcidA{} and Tam{\'a}s Cs{\"o}rg\H{o} 
 $^{1,2}$ \orcidB{}}

\AuthorNames{G{\'a}bor Kasza  and Tam{\'a}s Cs{\"o}rg\H{o}}

\AuthorCitation{Kasza, G.; Cs{\"o}rg\H{o}, T.}

\address{%
$^{1}$ \quad HUN-REN Wigner RCP, P.O. Box 49, H-1525 Budapest, Hungary; tcsorgo@cern.ch 
\\
$^{2}$ \quad MATE Institute of Technology KRC, H-3200 Gy{\"o}ngy{\"o}s, Mátrai út 36, 
 Hungary}

\corres{Correspondence: kasza.gabor@wigner.hun-ren.hu}

\abstract{From a recently found family of analytic, finite and accelerating 1+1-dimensional solutions to perfect fluid relativistic hydrodynamics, we derive simple and powerful formulae to describe the rapidity and pseudorapidity density distributions. By introducing a new scaling function, we notice that the rapidity distribution data of the different experiments all collapse into a single curve. This data-collapsing (or scaling) behaviour in the rapidity distributions suggests that high-energy $p+p$ collisions may be described as collective systems.}


\begin{document}

\section{Introduction}

It is experimentally known that, when very fast particles collide, a~large number of new particles are created. The~study of the process of high-energy collisions based on thermodynamic methods was first proposed by Fermi~\cite{Fermi:1950jd,Fermi:1951zz}. However, Fermi's theory proved to be incomplete, which, in short, was due to his incorrect view of the expansion of the complex system~\cite{Belenkij:1955pgn}. As Landau has shown, the~expansion of the system can be studied using relativistic hydrodynamics~\cite{Landau:1953wku}.

In ref.~\cite{Landau:1953wku}, Landau has discussed in detail that, at the moment that two nucleons collide, a~large number of particles are created and the mean free path of the collisions between the particles is small compared to the size of the system; thus,~the medium reaches statistical equilibrium. The~expansion of the medium should be studied in a hydrodynamic framework as a perfect fluid, since the mean free path during the expansion process remains small compared to the size of the system. Due to the presence of huge energy density, particles are continuously created and absorbed in the medium, so the system can hardly be characterised by the number of particles~\cite{Belenkij:1955pgn}. As the system expands, the~interaction between particles decreases, so the mean free path increases. When the interaction becomes sufficiently small, the~number of particles can be considered a physical property. When the mean free path becomes comparable to the size of the system, the~latter disintegrates into separate particles~\cite{Belenkij:1955pgn}.

The Hwa--Bjorken solution to relativistic hydrodynamics is suitable for describing \mbox{1+1}-dimensional explosive perfect fluids~\cite{Hwa:1974gn,Bjorken:1982qr}. The solution utilizes a Hubble-type velocity field and one of its main shortcomings is that it assumes zero relativistic acceleration. Consequently, it predicts a rapidity plateau, or~a flat rapidity distribution, which does not describe any of the high-energy experimental data at RHIC or LHC energies (except, perhaps, in a narrow interval near midrapidity, where the distribution is approximately constant near its maximum). 
However, one of the great advantages of this Hwa--Bjorken solution is that it uses simple formulae, and,~from this solution, Bjorken derived his famous formula for estimating the initial energy density of the hot and dense matter created in heavy-ion collisions~\cite{Bjorken:1982qr}. This solution has previously been generalised to the case of an accelerating velocity field~\cite{Csorgo:2006ax,Nagy:2007xn}, but~these generalisations only satisfy the equations of hydrodynamics under strong constraints. In~one of these new exact solutions~\cite{Csorgo:2006ax,Nagy:2007xn}, the rate of acceleration may have an  arbitrary constant value, but~only for a super-hard, non-realistic equation of state, where the speed of sound and the speed of light are equal. Recently, Csörgő, Kasza, Csanád and Jiang published a 1+1-dimensional generalisation of the Hwa--Bjorken solution, where the rate of acceleration of the velocity field may have an arbitrary constant value, even for a realistic equation of state~\cite{Csorgo:2018pxh}, referred to as the CKCJ~solution.

This CKCJ family of perfect fluid solutions and its applications have been published previously. The~calculation of the pseudorapidity distribution and the fitting of the resulting formula to experimental data were published in refs.~\cite{Csorgo:2018pxh,Csorgo:2018fbz,Kasza:2018qah}. The~correction to Bjorken's initial energy density estimate was presented in detail in ref.~\cite{Kasza:2018jtu}. Based on measured data, we have determined the lifetime parameter of heavy-ion collisions for $\sqrt{s_{NN}}=130$ and 200~GeV $Au+Au$ reactions and published the result in refs.~\cite{Kasza:2018qah,Csorgo:2018crb}. We calculated the initial energy density in $\sqrt{s_{NN}}=130$ and 200 GeV $Au+Au$ collisions, which were also published in~\cite{Kasza:2018qah}. The~most recent result is the evaluation of the thermal photon spectrum, which we compared with experimental data in ref.~\cite{Kasza:2023rpx}. 

{Similar efforts were carried out by Csanád, Májer and Vargyas: they evaluated the direct photon spectrum~\cite{Csanad:2011jr,Csanad:2011jq} and hadronic observables~\cite{Csanad:2009wc} from a 1+3-dimensional analytic, but~boost-invariant, solution to relativistic hydrodynamics for perfect fluids. Recently, a \mbox{1+1}-dimensional perturbative solution to viscous hydrodynamics was used to derive an analytical formula for the rapidity and pseudorapidity distributions, analytically providing the correction of the bulk viscosity effect~\cite{Jiang:2018qxd}. The shear and bulk viscous hydrodynamic deformation of the rapidity distributions were also investigated by Hirano and Monnai, using the Color Glass Condensate approach for relativistic heavy-ion collisions~\cite{Monnai:2011ju,Monnai:2011pb,Monnai:2011zc}}.

The justification for the use of a perfect fluid solution has been established in some of our previous works: we have found the property of asymptotic perfection for both nonrelativistic~\cite{Kasza:2022spy} and relativistic~\cite{Csorgo:2020iug} Hubble-type viscous solutions. This asymptotic perfection means that viscous solutions become asymptotically equal with corresponding perfect fluid solutions, which implies that the results of the final state measurements of hadronic observables can be described not only by viscous solutions but also by the corresponding perfect fluid~solutions.

In this manuscript, using a new formula computed from the solution reported in ref.~\cite{Csorgo:2018pxh}, we find a data-collapsing or -scaling behaviour in experimental rapidity distribution data, which is one of the manifestations of a beautiful hydrodynamic scaling behaviour. {This phenomenon was observed by Carruthers and Duong-van at intermediate energies for $p+p\rightarrow \pi^+ + X$ and $p+p\rightarrow \pi^- + X$ reactions in refs.~\cite{Carruthers:1972pwk,Carruthers:1973ws}. They introduced a Gaussian scaling function for the rapidity densities from Landau hydrodynamics. In~this paper, we derive the same scaling function from a different hydrodynamic solution, and~we show that the data-collapsing behaviour in rapidity densities holds for high-energy collisions.} Our theoretical insight is successfully cross-checked against experimental data, as~shown explicitly in Section~\ref{sec:scaling}.

\section{Recapitulation of the CKCJ~Solution}
In this section, we briefly recapitulate the 1+1-dimensional, analytical, accelerating family of CKCJ solutions to relativistic perfect fluid hydrodynamics from ref.~\cite{Csorgo:2018pxh}. The~solutions of this family solve the equations of hydrodynamics at zero chemical potential ($\mu=0$) under the following equation of state:
\begin{equation}
    \varepsilon = \kappa_0 p,
\end{equation}
where $\varepsilon$ stands for the energy density, while the pressure is denoted by $p$ and $\kappa_0$ stands for their average (temperature-independent) constant of proportionality. This quantity $ \kappa_0$ is directly related to the speed of sound in the medium ($\kappa_0=c_s^{-2}$), as~recently clarified in ref.~\cite{Kasza:2022spy}.
For the four-velocity ($u^{\mu}$), 
the~CKCJ generalisation of the Hwa--Bjorken flow that allows the description of accelerating fluids reads as follows~\cite{Csorgo:2018pxh}:
\begin{equation}
    u^{\mu}=\left(\cosh\left(\Omega\right),\sinh\left(\Omega\right)\right),
\end{equation}
where $\Omega$ is the fluid rapidity, which is a function of the coordinate rapidity $\eta_z$, but~it is independent from the longitudinal proper time $\tau$ (i.e., $\Omega\equiv \Omega(\eta_z)$) \cite{Csorgo:2018pxh}. The new family of CKCJ solutions~\cite{Csorgo:2018pxh} can be summarised as follows:
\begin{eqnarray}
	\eta_z(H)  & = & \Omega(H) -H, \\ 
	\Omega(H)  & = & \frac{\lambda}{\sqrt{\lambda-1}\sqrt{\kappa_0-\lambda}}
	\textnormal{arctan}\left(\sqrt{\frac{\kappa_0-\lambda}{\lambda-1}}\textnormal{tanh}\left(H\right)\right), \\
	\sigma(\tau,H) & = &
	\sigma_0 \left(\frac{\tau_0}{\tau}\right)^{\lambda}\mathcal{V}(s)
		\left[1+\frac{\kappa_0-1}{\lambda-1}\textnormal{sinh}^2(H)\right]^{-\frac{\lambda}{2}},\\
	T(\tau,H) & = &T_0 
	\left(\frac{\tau_0}{\tau}\right)^{\frac{\lambda}{\kappa_0}} 
	\frac{1}{\mathcal{V}(s)}
		\left[1+\frac{\kappa_0-1}{\lambda-1}\textnormal{sinh}^2(H)\right]^{-\frac{\lambda}{2\kappa_0}},\label{eq:CKCJ_temp}\\
	s(\tau,H) & = & \left(\frac{\tau_0}{\tau}\right)^{\lambda-1} 
		\textnormal{sinh}(H)\left[1 + \frac{\kappa_0-1}{\lambda-1}\textnormal{sinh}^2(H)\right]^{-\lambda/2},
\end{eqnarray}
where $T$ is the temperature and $\sigma$ stands for the entropy density. The~trajectory equation of the fluid element is defined by the scaling variable $s$, satisfying the scale equation $u^{\mu}\partial_{\mu}s=0$. The~function $\mathcal{V}(s)$ is a non-negative function of the scaling variable $s$. The~parameter $\lambda$ is an integration constant, but~it has an important physical meaning, since it determines the acceleration of the velocity field: for $\lambda > 1$, we talk about accelerating expansion, while, for $0< \lambda < 1$, we consider decelerating expansion~\cite{Csorgo:2018pxh}.

The above equations can be considered as a parametric family of solutions in the sense that the coordinate-rapidity dependence of the thermodynamic quantities ($T$, $\sigma$, $p$, $\varepsilon$) and the elements of the velocity field ($\Omega$, $v_z$) are given through the parameter $H=\Omega(\eta_z)-\eta_z$~\cite{Csorgo:2018pxh}.

It is important to note that the above equations define a finite family of solutions in terms of the coordinate rapidity, and~the range of validity of the solutions is affected by the values of $\lambda$ and $\kappa_0$. However, in~the accelerationless limiting case $\lambda\rightarrow 1$, the~width of the range of validity goes to infinity~\cite{Csorgo:2018pxh}. Thus, for low accelerations, this family of solutions can be applied in the physical domain of high-energy collisions. If~the scaling function $\mathcal{V}(s)$ is set to $\mathcal{V}(s)=1$, then the Hwa--Bjorken solution can be recovered in the $\lambda\rightarrow 1$ limit:
\begin{eqnarray}
	\Omega  & = & \eta_z, \\ 
	\sigma(\tau) & = &
	\sigma_0 \left(\frac{\tau_0}{\tau}\right)^{\lambda},\\
	T(\tau) & = &T_0 
	\left(\frac{\tau_0}{\tau}\right)^{\frac{\lambda}{\kappa_0}}.
\end{eqnarray}

\section{Evaluation of the Rapidity~Density}
The Cooper--Frye formula can be used to calculate the rapidity distribution:
\begin{equation}\label{eq:Cooper-Frye-formula}
    \frac{dN}{dy}=\frac{1}{2\pi \hbar}\int d\Sigma_{\mu} p^{\mu} \exp\left(-\frac{p_{\mu}u^{\mu}}{T_{\rm F}\left(\tau_{\rm f},\eta_z\right)}\right),
\end{equation}
where $d\Sigma_{\mu}$ is the normal vector of the freeze-out hypersurface, while $p^{\mu}$ is the four-momentum of the detected particles. The~integral is performed on the freeze-out hypersurface, which is the set of points in spacetime $\left(\tau,\eta_z \right)$ where the hadronic medium is freezing out. This integral is evaluated at vanishing chemical potential; thus, the fugacity yields a trivial factor of unity in the Boltzmann integral, where $T_{\rm F}$ is the temperature of the freeze-out hypersurface, which is not a constant, since the points on the freeze-out hypersurface have different temperatures. The~proper time associated with the freeze-out is denoted by $\tau_{\rm f}$. For~the temperature, we use Equation~\eqref{eq:CKCJ_temp}, and~we consider the simplest possible case, i.e.,~when the scaling function is chosen to be unity: $\mathcal{V}(s)=1$.

The new solution was found in 1+1-dimensional spacetime, so it was necessary to embed the formula for the rapidity distribution in 1+3-dimensional spacetime. To~accomplish this, we first assumed that the temperature of the medium is homogeneous in the transverse plane, and,~thus, the rapidity density is independent of the transverse coordinates. As~a second step, we introduced the transverse mass $m_{\rm T}$ instead of the particle mass $m$. In~such a case, the~integration of the distribution on the surface perpendicular to the beam direction can be performed trivially, but~the integration on momentum space must also be performed. Turning to the transverse mass as an integration variable, the~rapidity distribution embedded in the three-dimensional space can be obtained by integrating the double-differential spectrum over the transverse mass.  We used saddle point approximation in the calculations, which led to the following result:
\begin{equation}\label{eq:CKCJ_rapiditydist_1 + 3_Tf}
\frac{dN}{dy}\approx\left.\frac{dN}{dy}\right|_{y=0}\cosh^{-\frac{\alpha(\kappa_0)}{2}-1}\left(\frac{y}{\alpha}\right)\exp\left(-\frac{m}{T_{\rm f}}\left[\cosh^{\alpha(\kappa_0)}\left(\frac{y}{\alpha}\right)-1\right]\right),
\end{equation}
where $\alpha(\kappa_0)=(2\lambda-\kappa_0)/(\lambda-\kappa)$ and~$\alpha(1)=\alpha$. The~kinetic freeze-out temperature is denoted by $T_{\rm f}$ and $y$ stands for the~rapidity.

In Equation~\eqref{eq:CKCJ_rapiditydist_1 + 3_Tf}, the contribution of radial flow is not present, but,~from the $p_{\rm T}$-spectra, it is known that the radial flow effect is not negligible. While the embedding of the rapidity distribution in three spatial dimensions is an efficient tool, it makes the assumption that the temperature is homogeneous in the transverse plane. For~this reason, the~effect of radial flow does not appear, so we artificially include it in the equations. We introduce the effective temperature $T_{\rm eff}$, which is equal to the sum of the kinetic freeze-out temperature $T_{\rm f}$ and the contribution of the radial flow. In~the formula for the rapidity distribution, we simply use the notation $T_{\rm f} \rightarrow T_{\rm eff}$ to include the radial flow contribution into the distribution~\cite{Csorgo:2018pxh,Csorgo:2018fbz}:
\begin{equation}\label{eq:CKCJ_rapiditydist_1 + 3}
\frac{dN}{dy}\approx\left.\frac{dN}{dy}\right|_{y=0}\cosh^{-\frac{\alpha(\kappa_0)}{2}-1}\left(\frac{y}{\alpha}\right)\exp\left(-\frac{m}{T_{\rm eff}}\left[\cosh^{\alpha(\kappa_0)}\left(\frac{y}{\alpha}\right)-1\right]\right).
\end{equation}
The midrapidity density can be expressed by the following formulae:
\begin{align}
    \left.\frac{dN}{dy}\right|_{y=0}&=\frac{1}{\sqrt{\lambda(2\lambda-1)}}\frac{V_xV_p}{(2\pi\hbar)^3}\exp\left(-\frac{m}{T_{\rm eff}}\right),\\
    V_x&=R^2\pi\tau_{\rm f},\\
    V_p&=(2\pi m T_{\rm eff})^{3/2}.
\end{align}
Here, $R^2\pi$ is the finite size of the transverse plane, $V_x$ is the volume of the coordinate space and~$V_p$ is the volume of the momentum space. While the midrapidity density depends on several parameters of the model, for~simplicity, we now consider it only as a normalisation constant, which will greatly facilitate the fitting of experimental~data. 

\subsection{The Approximate Formula for the Rapidity~Distribution}
We now summarise, in a few lines, that further approximations can be made on Equation~\eqref{eq:CKCJ_rapiditydist_1 + 3} of the rapidity distribution in the region where $|y| \ll \alpha = 2 + 1/(\lambda-1)$. Obviously, this region widens rapidly if one approaches the accelerationless limiting case ($\lambda \rightarrow 1$). In~this approximation, the~rapidity distribution becomes a Gaussian distribution~\cite{Kasza:2018qah}:
\begin{equation}\label{eq:CKCJ_gaussian_dNdy}
    \frac{dN}{dy} \approx \frac{\langle N \rangle}{\left(2\pi\Delta y ^2\right)^{1/2}}\exp\left(-\frac{y^2}{2\Delta y^2}\right).
\end{equation}

In this approximation, the~two new parameters are the average multiplicity $\langle N \rangle$ and $\Delta y$, which characterises the width of the distribution. Both quantities can be expressed in terms of the parameters of the rapidity density embedded in the three-dimensional space~\cite{Kasza:2018qah}:
\begin{align}
    \frac{1}{\Delta y^2} &= \left(\lambda-1\right)^2 \left[ 1 + \left(1+\frac{1}{\kappa_0}\right)\left(\frac{1}{2}+\frac{m}{T_{\rm eff}}\right) \right],\label{eq:CKCJ_gaussian_width_y} \\
    \langle N \rangle &= \left(2\pi\Delta y ^2\right)^{1/2} \left.\frac{dN}{dy}\right|_{y=0} \label{eq:CKCJ_mean_multiplicity}.
\end{align}

{Such a Gaussian rapidity distribution is by no means a unique result that is characteristic only to the CKCJ solution. Other calculations and  model results that are based on Landau hydrodynamics usually predict a similar result; see, e.g.,~refs.~\cite{Belenkij:1955pgn,Carruthers:1972pwk,Carruthers:1973ws,Shuryak:1972zq,Zhirov:1975qu,Steinberg:2004vy,Wong:2008ex,Wong:2014sda,Sen:2015hwa,Gazdzicki:2010iv,Jiang:2013rm}. It is shown in ref.~\cite{Rahman:2021rxl} that Landau hydrodynamics can describe both the measured rapidity distributions and the results of the EPOS 1.99 event generator. Recently, the~Landau hydrodynamical model was used in combination with an ANN (Artifical Neural Network) simulation model to estimate the $dN/dy$ distributions in central $Au+Au$ collisions for two future experimental facilites: FAIR and NICA~\cite{Habashy:2021poi}. }

{
Despite the similarity of the Gaussian rapidity distributions in the Landau and in the CKCJ solutions, there is an important technical difference:
the Landau solution is an implicit solution, where the time and coordinate variables are given in terms of the temperature and the fluid rapidity, while the CKCJ solution is a parametric, but~explicit, solution, where the observables are expressed in terms of the acceleration parameter $\lambda$, the~average speed of sound 
$\kappa_0$, the~effective temperature (slope parameter) at midrapidity $T_{\rm eff}$ and the midrapidity density $\left. dN/dy\right|_{y=0}$---see Equations~\eqref{eq:CKCJ_rapiditydist_1 + 3} and \eqref{eq:CKCJ_gaussian_dNdy}. This becomes advantageous when analysing the scaling properties of the rapidity distribution, as~it gives a beautiful example of hydrodynmamical scaling.
}

Using Equation~\eqref{eq:CKCJ_gaussian_dNdy}, the~experimental data can be described by fitting two parameters ($\langle N \rangle$ and $\Delta y$), while the original formula (Formula 
\eqref{eq:CKCJ_rapiditydist_1 + 3})
 has four parameters to be fitted ($\kappa_0$, $\lambda$, $T_{\rm eff}$, 
$\left. dN/dy\right|_{y=0}$).
This result is a wonderful manifestation of hydrodynamic scaling behaviour: two different fits of the data, resulting in different parameter sets, can lead to the same curve, provided that the values of the two relevant combinations of the four original parameters, \eqref{eq:CKCJ_gaussian_width_y} and \eqref{eq:CKCJ_mean_multiplicity}, remain the same. From~Equation~\eqref{eq:CKCJ_gaussian_dNdy} of the rapidity distribution, it is also clear that the rapidity density can be normalised by both the midrapidity value of the distribution and the mean~multiplicity.

\subsection{The Approximate Formula for the Pseudorapidity~Distribution}

The pseudorapidity distribution can be calculated from the double-differential invariant momentum spectrum and can be written as the product of two factors (as detailed in ref.~\cite{Kasza:2018qah}). One factor is the rapidity density, as~written in Equation~\eqref{eq:CKCJ_rapiditydist_1 + 3}. The~other factor is the Jacobian determinant $\mathcal{J}$, calculated for a mean rapidity-dependent transverse momentum: $\langle p_{\rm T}(y) \rangle$~\cite{Kasza:2018qah}. The~results in refs.~\cite{EHSNA22:1997ddc,Csorgo:2004id}  have shown that the effect of transverse flow goes through the rapidity dependence of the average transverse momentum dependence of $\mathcal{J}$. {The rapidity-dependent average transverse mass was evaluated by saddle-point approximation, which limits our investigation to the nearly exponential ($p_T < 1-2$ GeV) kinematic limit. In~the leading order, the~rapidity dependence of the average transverse mass has Lorentzian shape~\cite{Csorgo:2018pxh,Kasza:2018qah}}:
\begin{equation}\label{eq:average_mT}
    \langle m_{\rm T}(y) \rangle = \sqrt{\langle p_{\rm T}(y)\rangle^2+m^2} = m+\frac{T_{\rm eff}}{1+by^2},
\end{equation}
where $b=\alpha(\kappa_0)/(2\alpha^2)$ controls the width of the Lorentz distribution. 
{This result was mentioned first just after Equation~(36) in ref.~\cite{Csorgo:2018pxh} and it has been discussed, in more detail, that the CKCJ solution has a realistic feature: a~rapidity-dependent decrease in the average transverse momentum. According to Equation~\eqref{eq:average_mT},} the physical meaning of the effective temperature becomes clear: it is related to the thermally averaged transverse mass at midrapidity. A~Lorentzian shape for the rapidity-dependence of the average transverse momentum has been observed by the EHS/NA22 experiment in slightly asymmetric hadron--proton collisions~\cite{EHSNA22:1997ddc}.

The pseudorapidity distribution and the partial results of its derivation were described in detail in ref.~\cite{Kasza:2018qah}, so we leave it aside for now. In~this manuscript, we focus on an approximate formula that holds in the limiting case $\lambda \rightarrow 1$. Note that, in this limit, $b\rightarrow 0$, so the average rapidity-dependent transverse momentum can be approximated by its central value: $\langle p_{\rm T}(y)\rangle \rightarrow \langle p_{\rm T}(0) \rangle = \langle p_{\rm T} \rangle = \sqrt{T_{\rm eff}^2+2mT_{\rm eff}}$. In~this case, the~Jacobi determinant and the pseudorapidity distribution can be approximated using the following equations~\cite{Kasza:2018qah}:
\begin{align}
    \mathcal{J} &\approx \frac{\cosh(\eta_{\rm p})}{\sqrt{D^2+\cosh^2(\eta_{\rm p})}}, \label{eq:CKCJ_gaussian_Jacobi}\\
    \frac{dN}{d\eta_{\rm p}} &\approx \frac{\langle N \rangle}{\sqrt{2\pi \Delta y^2}} \frac{\cosh(\eta_{\rm p})}{\sqrt{D^2+\cosh^2(\eta_{\rm p})}} \exp \left(-\frac{y^2\left(\eta_{\rm p}\right)}{2\Delta y^2}\right),\label{eq:CKCJ_gaussian_dNdetap}
\end{align}
where $\eta_p$ is the pseudorapidity and~$D=m/\langle p_{\rm T}\rangle$ stands for the depthness parameter. In~Equation~\eqref{eq:CKCJ_gaussian_dNdetap}, the~$y(\eta_{\rm p})$  rapidity can be approximated as follows~\cite{Kasza:2018qah}:
\begin{equation}
    y(\eta_{\rm p}) \approx \tanh^{-1} \left(\frac{\cosh(\eta_{\rm p})}{\sqrt{D^2+\cosh^2(\eta_{\rm p})}} \tanh\left(\eta_{\rm p}\right)\right).
\end{equation}

\section{The Scaling of \boldmath$dN/dy$ Data}
 \label{sec:scaling}

The Gaussian density is expressed by Equation~\eqref{eq:CKCJ_gaussian_dNdy} and is determined by the normalisation factor and the width of the distribution. These parameters thus carry the characteristics of the different reactions, such as the collision energy, the~size of the colliding system and the centrality of the collision. The~normalisation factor and the width of the distribution can be scaled out from Equation~\eqref{eq:CKCJ_gaussian_dNdy}, and,~with that, we obtain a curve that is independent from the characteristics of the reactions. We introduce the variable $x$, which is the ratio of the rapidity $y$ to the width $\Delta y$:
\begin{equation}
    x=\frac{y}{\Delta y}.
\end{equation}

Then, we divide the rapidity distribution by the normalisation factor, introducing a scale-independent distribution:
\begin{equation}\label{eq:CKCJ_scaled_dNdy}
    \left(\left.\frac{dN}{dy}\right|_{y=0}\right)^{-1}\frac{dN}{dy} = \exp\left(-\frac{x^2}{2}\right),
\end{equation}
where the expression of the normalisation factor is given by Equation~\eqref{eq:CKCJ_mean_multiplicity}, which is valid in the $|y|\ll \alpha$ range. Our idea is that, if we scale back the rapidity distribution data from different reactions with the normalisation factor and $\Delta y$ as above, then, regardless of the reaction, all the data series will follow the curve of Equation~\eqref{eq:CKCJ_scaled_dNdy}. However, in~most cases, we do not have $dN/dy$ measurements available, so, in order to check our conjecture, we have transformed the available pseudorapidity density data. For~this reason, first we used Equation~\eqref{eq:CKCJ_gaussian_dNdetap} to determine the parameters $D$, $\langle N \rangle$ and $\Delta y$ from the pseudorapidity distribution data. Thus, for~each dataset, we obtained different values of $D$, $\langle N \rangle$ and $\Delta y$. Using the $D$ parameter, we extracted the Jacobian determinant to transform the pseudorapidity distribution data points into rapidity distribution data points. Then the mean multiplicity $\langle N \rangle$ and the width parameter $\Delta y$ were used to rescale the rapidity distributions. Using this method, we checked the manifestation of the scaling behaviour on 11 datasets: PHOBOS 0-30\% $Au+Au$ at $\sqrt{s_{NN}}=20$, 62.4, 130~\cite{PHOBOS:2010eyu},  CMS $p+p$ at $\sqrt{s}=7$, 8 and 13 TeV and 0-80\% $Xe+Xe$ at $\sqrt{s_{NN}}=5.44$ TeV~\cite{CMS:2010tjh,CMS:2014kix,CMS:2017eoq,CMS:2019gzk} and~ALICE $Pb+Pb$ at $\sqrt{s_{NN}}=5.02$ TeV for central, mid-central and peripheral collisions~\cite{ALICE:2016fbt}. We note that the PHOBOS data
required a special approach, as~PHOBOS did not publish the statistical and systematic errors of the pseudorapidity distributions separately. For~this reason, we used the best estimate for the statistical errors in the PHOBOS data, described in detail in ref.~\cite{Kasza:2018qah}.

\begin{figure}[H]
      \includegraphics[scale=0.9]{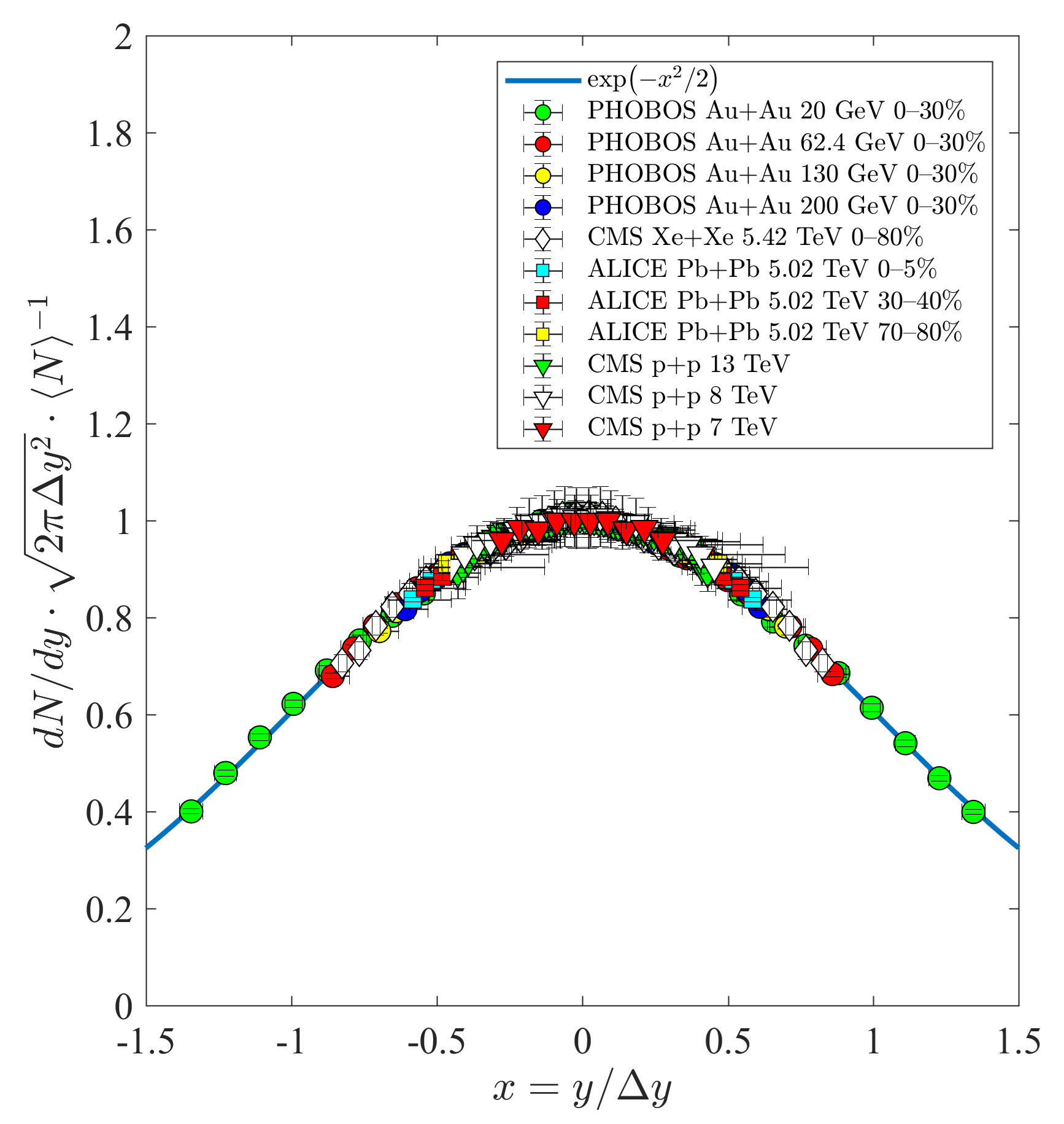}
    \caption{Manifestation of hydrodynamic scaling behaviour on experimental data: by transforming the pseudorapidity distributions measured in different reactions, we calculated the data points associated with the rapidity densities, which we rescaled by the normalisation factor. The~width of the distributions were also scaled by introducing the variable $x$. All of the resulting data points lie within error on the scale-independent blue curve obtained from our theoretical calculations. The~datasets were taken from CMS~\cite{CMS:2010tjh,CMS:2014kix,CMS:2017eoq,CMS:2019gzk}, ALICE~\cite{ALICE:2016fbt} and PHOBOS~\cite{PHOBOS:2010eyu}.}
    \label{fig:CKCJ_dNdy_hydroscaling}
\end{figure}

A comparison of these data with Equation~\eqref{eq:CKCJ_scaled_dNdy} is shown in Figure~\ref{fig:CKCJ_dNdy_hydroscaling}. The~vertical errors of the data points correspond to the quadrature of the statistical errors associated with the pseudorapidity distributions and the uncertainty in the Jacobian determinants. The~horizontal error of the data points is due to the uncertainty in $\Delta y$. Figure~\ref{fig:CKCJ_dNdy_hydroscaling} convincingly illustrates that our conjecture was right, i.e.,~all the data points measured in the different reactions are clustered (within error) on the blue curve, which is the manifestation of hydrodynamic scaling behaviour in the experimental~data.

\begin{figure}[H]
    \centering
    \includegraphics[scale=0.9]{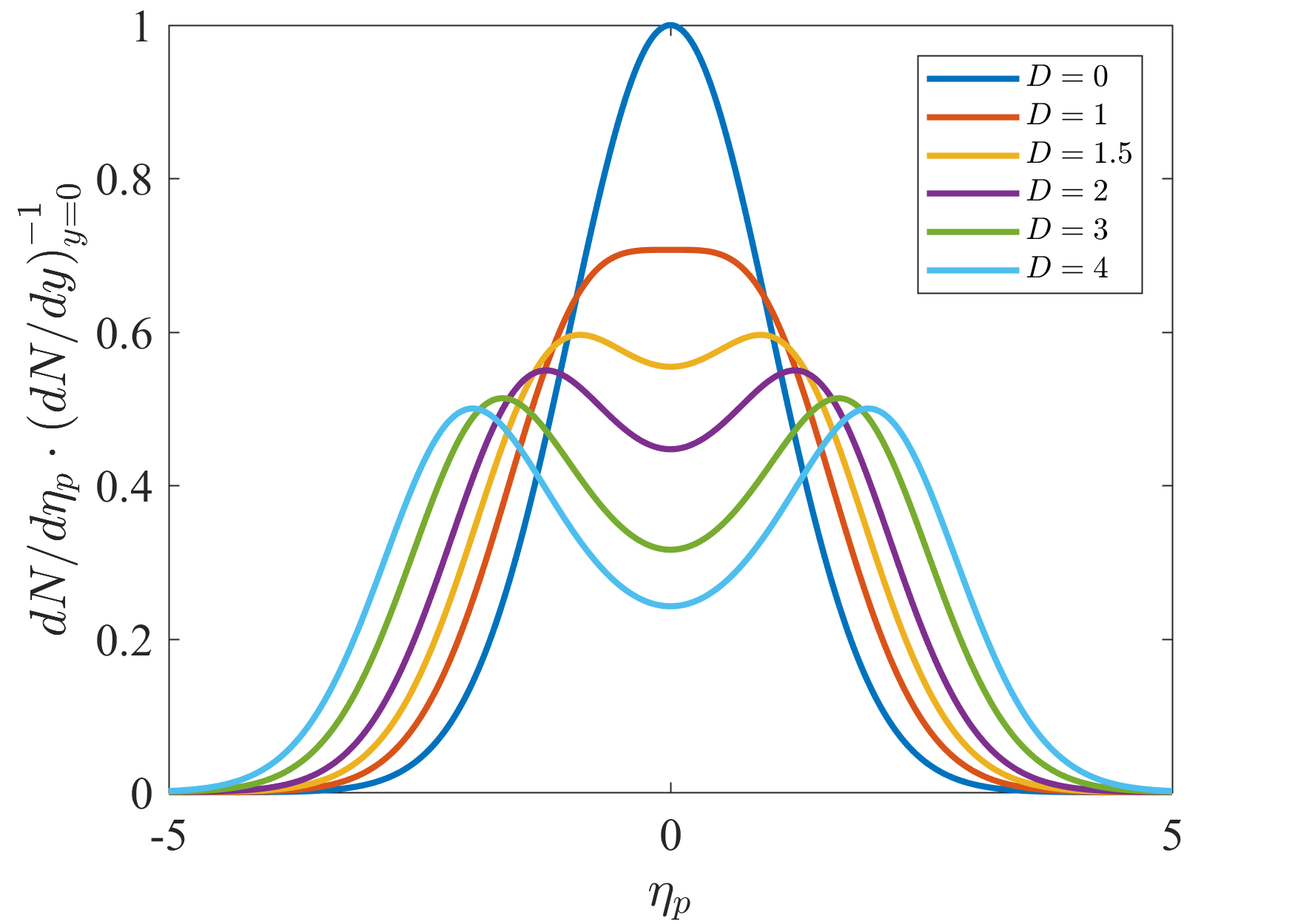}
    \caption{The pseudorapidity distribution (divided by $\left.dN/dy\right|_{y=0}$) for different $D$ values. The~width parameter $\Delta y$ is fixed to 1. It can be clearly seen that, as $D$ approaches 0 more closely, midrapidity dip in the pseudorapidity distribution becomes smaller, and,~for $D=0$, we recover Equation~\eqref{eq:CKCJ_gaussian_dNdy} for the shape of the rapidity~distribution.}
    \label{fig:CKCJ_dNdeta_nonscaling}
\end{figure}

For the pseudorapidity distributions, we could not find a scale function similar to Equation~\eqref{eq:CKCJ_scaled_dNdy} to describe the collapse of the experimental datasets of $dN/d\eta_{\rm p}$, as~we saw for the rapidity distributions in Figure~\ref{fig:CKCJ_dNdy_hydroscaling}. This is due to the presence of the depth parameter $D$, which controls the dip of the pseudorapidity distribution around midrapidity. This can be easily seen by evaluating Equation~\eqref{eq:CKCJ_gaussian_dNdetap} for the case of zero pseudorapidity:
\begin{equation}
    \left.\frac{dN}{d\eta_{\rm p}}\right|_{\eta_{\rm p}=0}=\left.\frac{dN}{dy}\right|_{y=0}\frac{1}{\sqrt{1+D^2}},
\end{equation}
where we used Equation~\eqref{eq:CKCJ_mean_multiplicity} for the midrapidity density. Thus, if~this dip is close to zero ($D\approx 0$), we recover the formula of the rapidity distribution and the scaling behaviour is restored. This phenomenon is illustrated in Figure~\ref{fig:CKCJ_dNdeta_nonscaling}.

\unskip

\section{Conclusions}
In this manuscript, we introduced a new approximate formula to describe the rapidity distributions, based on the solution published in ref.~\cite{Csorgo:2018pxh}. By~rescaling the new formula, we introduced a new scale function to show the data collapsing in the rapidity distribution measurements. This data-collapsing (or scaling) behaviour was tested on 11 datasets from three different experiments (CMS, PHOBOS, ALICE). It is important to emphasise that these datasets are not only from nucleus--nucleus collisions, but~also from proton--proton collisions, which are also described by the curve of the scale function. From~this result, we can conclude that $p+p$ collisions can also be treated as  collective systems, as~far as their rapidity or pseudorapidity distributions are concerned. Let us mention that the $p+p$ data can be well-described with low speed of sound values ($c_s^2 = 0.1$), which is typical for fluids. This suggests that a strongly interacting quark--gluon plasma may be formed in $p+p$ collisions at RHIC and at LHC energies. Although~$p+p$ collisions at RHIC energies have not been investigated in this manuscript, it is expected that the rapidity distributions of these systems do not violate the scaling behaviour presented in Section~\ref{sec:scaling}. It has been shown, in ref.~\cite{Csorgo:2004id}, that $p_{\rm T}$ spectra and HBT radii measured in $\sqrt{s}=200$ GeV $p+p$ collisions at RHIC can be described by the Buda--Lund hydrodynamical~model.

Let us emphasise that the hydrodynamical description of small systems (hadron--hadron collisions) is not a new idea. The~EHS/NA22 collaboration  previously estimated some hydrodynamic model parameters (radial flow, kinetic freeze-out temperature, geometric radius of the particle source, mean lifetime) via invariant momentum distributions of $\pi^-$ particles and Bose--Einstein correlation functions in $(\pi^+/K^+)p$ reactions at $\sqrt{s}\approx 22 $ GeV, where the mean multiplicity of charged particles was less then 10~\cite{EHSNA22:1997ddc}.

\vspace{6pt}
\authorcontributions{Writing---original draft preparation, G.K.; writing---review and editing, T.Cs. and G.K.; conceptualisation, T.Cs.; supervision, T.Cs.; visualisation, G.K. All authors have read and agreed to the published version of the manuscript.}  

\funding{This research was funded by NKFIH (National Office of Research, Innovation and Development, Hungary) K-133046 and K-147557, and the KKP-2023 Research Excellence Programme of MATE, Hungary. The APC was funded by NKFIH K-133046.} 

\dataavailability{No new data were created or analyzed in this study. Data sharing is not applicable to this article.} 

\conflictsofinterest{The authors declare no conflicts of interest.}  


\reftitle{References}

\begin{adjustwidth}{-\extralength}{0cm}

\PublishersNote{}
\end{adjustwidth}




\end{document}